\documentclass[12pt]{revtex4}
\usepackage{graphicx}
\usepackage{psfig}
\newcommand{\be}{\begin{equation}}
\newcommand{\bea}{\begin{eqnarray}}
\newcommand{\ee}{\end{equation}}
\newcommand{\eea}{\end{eqnarray}}

\begin{document}
\title{Vortex creation during magnetic trap manipulations of
spinor Bose-Einstein condensates}

\author{A.P. Itin,$^{1,2}$ T. Morishita,$^1$
M. Satoh,$^1$ O.I. Tolstikhin,$^3$ and S. Watanabe$^1$}
\affiliation{$^1$Department of Applied Physics and Chemistry,
University of Electro-Communications,\\
1-5-1, Chofu-ga-oka, Chofu-shi, Tokyo 182-8585, Japan\\
$^2$Space Research Institute, RAS, Moscow, Russia\\
$^3$Russian Research Center ``Kurchatov Institute'',
Kurchatov Square 1, Moscow 123182, Russia}

\begin{abstract}
We investigate several mechanisms of vortex creation during splitting
of a spinor BEC in a magnetic trap controlled by a pair of current
carrying wires and bias magnetic fields. Our study is motivated by a
recent MIT experiment on splitting BECs with a similar trap, where
unexpected fork-like structure appeared in the interference fringes
corresponding to interference of two condensates, one with and the
other without a singly quantized vortex. It is well-known that in a
spin-1 BEC in a quadrupole trap a doubly quantized vortex is produced
topologically by a ``slow'' reversal of bias magnetic field $B_z$.  We
find that in the magnetic trap considered it is also possible to
produce a 4- and 1-quantized vortex in a spin-1 BEC. The latter is
possible, for example, during the magnetic field switching-off process.
We therefore provide a possible explanation for the unexpected
interference patterns in the experiment. We also give an example of the
creation of singly quantized vortices due to  ``fast'' splitting, which
is a possible alternative mechanism of the interference pattern.
\end{abstract}
\maketitle
\newpage

\section{Introduction}

Coherent manipulation of matter-waves is presently a very
important experimental and theoretical field. Coherent splitting
of matter waves into spatially separate atomic wave packets with a
well-defined relative phase is necessary for atom interferometry
and quantum information processing applications. Bose-Einstein
condensates (BECs) open new unprecedented perspectives in this
direction. Coherent splitting of a BEC was recently realized both
in an optical double-well trap \cite{Shin2} and in a magnetic
chip-based double-well trap \cite{Shin}. In the latter experiment,
the magnetic trap was produced by a pair of current carrying wires
and a bias field used to control the distance between the wells
\cite{Hinds}. An intriguing feature of this experiment was
appearence of a fork-like structure in the absorption image of
interference fringes designating creation of a singly quantized
vortex in one of the wells. Thus perturbations during condensate
manipulations were violent enough to generate vortices. To
identify possible mechanisms of vortex creation is an inevitable
challenge for future pursuit of BEC interferometry with such kind
of an experimental configuration.

Here we study two independent scenarios that could lead to vortex
creation. One of them is the phase imprinting during the switching-off
process (due to different decay times of the bias fields and the fields
of the two wires), while the other is dynamical vortex creation during
fast splitting. Although the switching-off time in the particular
experiment \cite{Shin} was only about 20 $\mu$s (a duration much
shorter than the inverse of any trap frequency), it is slow as compared
to the Larmor frequency in the magnetic field of order of 1G.
Therefore, adiabatic imprinting might take place. On the other hand,
the authors of Ref. \cite{Shin} think the phase imprinting mechanism is
an unlikely explanation for the vortex creation since they have never
observed interference pattern of a doubly quantized vortex (which is
created topologically in spin-1 condensates when zero point of the
magnetic field crosses through condensates). We find realistic
scenarios for a singly quantized vortex interference pattern to appear
as a result of topological phase imprinting during the switching-off
process. Indeed as we assume exponential decay of magnetic fields
produced by the two wires and the bias field $B_z$ with different decay
constants, with parameters close to the experiment we obtain
numerically that a large part of the atoms (about half of the
condensate) can be transferred to the component $\Psi_0$
(i.e., with $m=0$ in $z$-quantized basis, see Eqs.~(\ref{zbasis1})
and (\ref{zbasis2}) for notation)
which has singly quantized vortices around each of the
two minima of the magnetic field. In contrast, when we try to obtain
doubly quantized vortices using bias field $B_z$ reversal, it turns out
to be more difficult. At such short timescales, only a small part of
the population is transferred to the component  $\Psi_1$ with doubly
quantized vortices, while the rest of the population is redistributed
among other components. Creation of doubly quantized vortices in such a
configuration requires times of the order of hundreds of $\mu$s. We
also present an example of dynamical vortex creation during fast
splitting. In this case,  the condensate (including the $\Psi_{-1}$
component) acquires winding phase during the splitting process. This
leads to clear fork-like structures in all components and thus the
total density upon expansion. Although in our calculations this
requires a sufficiently shorter time of splitting than that which was
actually effected in the real experiment \cite{Shin},
it might be caused by the
oversimplified model (without jitter or fluctuations of the magnetic
field, etc).

The rest of the paper is organized as follows. In Sec.~II we
introduce the model (spinor BEC described by 2D GP equation with
spin degrees of freedom), describe the configuration of the
magnetic field given by the two-wires setup, and its weak- and
strong-field seeking states. In Sec.~III, we present numerical
results on vortex creation in the system. In Sec.~IIIA we  study
several examples of how an unexpected topological phase imprinting
could have taken place in the MIT experiment. There we take into
account gravity and use realistic parameters relevant to the
experiment. We assume that after switching off the magnetic field
the $B_z$ field decays faster than transversal magnetic field
$B_{\perp}$, which can lead to the phase imprinting. In Sec.~IIIB
we present examples of dynamical vortex creation during splitting
process. When split slowly in 30 ms, the condensate develops no
vortex. However, reduction of the splitting duration to about 5
ms, vortices are created. Then, even in the case where the
switching-off process leaves all population in the initial
$\Psi_{-1}$ component, expansion of condensates produces very
clear forks in the interference pattern.

\section{The model}
\subsection{Spinor Bose condensate}

We consider a BEC of alkali atoms with hyperfine spin $F=1$ using the
$z$-quantized basis \cite{Machida}. The order parameter is expanded as
\be \Psi= \sum_{m=\pm 1,0} \Psi_m |m>, \label{zbasis1} \ee where $|m>$
are eigenvectors of $F_z$:
\be \label{zbasis2}
|1>= (1,0,0)^T, \quad |0> = (0,1,0)^T, \quad |-1> = (0,0,1)^T.
\ee
 We use the time-dependent form of the GP equation with spin
degrees of freedom developed in \cite{Machida}:
\bea i \frac{\partial}{\partial t} \Psi_j &=& \Bigl[  -\frac{\hbar^2 }{2 M}\nabla^2 +g_n\sum_l |\Psi_l|^2
+ V(\mathbf{r}) \Bigr]\Psi_j + \Bigl[ g_s \sum_{\alpha} \sum_{lp} (\Psi_l
(F_{\alpha})_{lp} \Psi_p)(F_{\alpha})_{jk}- {\cal B}_{jk} \Bigr] \Psi_k, \nonumber\\
{\cal B}_{jk} &=& \left(\begin{array}{ccc}
B_z & \frac{B_x - B_y}{\sqrt{2}} & 0 \\
 \frac{B_x + B_y}{\sqrt{2}} & 0  &  \frac{B_x - B_y}{\sqrt{2}}\\
 0 &  \frac{B_x + B_y}{\sqrt{2}} & -B_z,
\end{array}\right)
 \eea
where $F_{\alpha}$ $(\alpha=x,y,z)$ are the angular momentum operators
in the basis of the eigenvectors of $F_z$:

\bea F_x = \frac{1}{\sqrt{2}} \left(\begin{array}{ccc}
0 & 1 & 0 \\
1  & 0  & 1 \\
 0 & 1 & 0 \end{array}\right), \qquad
F_y = \frac{i}{\sqrt{2}} \left(\begin{array}{ccc}
0 & -1 & 0 \\
1  & 0  & -1 \\
 0 & 1 & 0 \end{array}\right), \qquad
 F_z = \frac{1}{\sqrt{2}} \left(\begin{array}{ccc}
 1  & 0  &  0 \\
 0  & 0  &  0 \\
 0  & 0  & -1 \end{array}\right), \nonumber
\eea and  $g_n= 4\pi \hbar^2 (a_0+2a_2)/3m$ is the spin-independent and
$g_s=4\pi \hbar^2 (a_2-a_0)/3m$ is the spin-dependent interaction
coefficients. Here the scattering lengths $a_0$ and $a_2$ characterize
collisions between atoms with total spin 0 and 2, $m$ is the atomic
mass.

We consider a condensate of  $^{23}$Na, which was used in the
experiment of \cite{Shin}. Parameters of Na are $M=3.81 \cdot 10^{-26}
$ kg, $a_2=2.75$nm, $a_0=2.46$nm \cite{Ketterle}. Spin-independent
potential $V(\textbf{r})$ is provided either by gravity as in the MIT
experiment
 ($V(\textbf{r})= - Mgy$, we consider this case in Section 3),  or, in principle,
``optical plug'' can be used. We assume the wavefunction of either
component does not depend on the $z$ coordinate, i.e. the condensate is
quasi-2D. We therefore use two-dimensional nonlinearity parameters
$g^{2D}_n$ and $g^{2D}_s$ which are related as $g^{2D}_n/g^{2D}_s =
27.44$. Configuration of the magnetic field is described in the next
subsection.

\subsection{The magnetic trap}

We use a two-wires setup suggested in Ref. \cite{Hinds}. It was used in
a recent experiment of Y. Shin \emph{et al.} \cite{Shin}, where several
important elements were added making the system essentially
three-dimensional. We consider a two-dimensional trap here to simplify
numerical calculations. Magnetic field produced by two parallel
current-carrying wires is given by

\bea B_x^W= \frac{-y}{(x+d)^2+ y^2} + \frac{-y}{(d-x)^2+y^2}  ,\nonumber\\
B_y^W= \frac{x+d}{(x+d)^2+ y^2} + \frac{-d+x}{(d-x)^2+y^2},
 \eea
where $2d$ is the distance between the wires.

 Additional  magnetic field $B_z$ is added along $z$ direction,
and bias field $B_x^B$ is added in $x$ direction: $B_x=B_x^W+ B_x^B$.
For weak-field-seeking atoms, amplitude of magnetic field $B(x,y)$
plays the role of trapping potential (while the potential for strong
field-seeking atoms is $-B(x,y)$). At certain critical $B_x^B =
B_{x_0}$ the trap potential $B(x,y)$ has a single minimum located on
distance $d$ away from the surface, at the middle of the two wires. At
$B_x^B$ greater or smaller than $B_{x_0} $ the potential has two wells:
when $\Delta B_x \equiv B_x^B - B_{x_0} > 0$, the two wells are
separated in the $x$ direction, while for $\Delta B_x < 0$ they are
separated in the $y$ direction and has equal $x$-coordinates of their
centers. Addition of bias magnetic field $B_y^B$ along the $y$-axis
rotates these two wells in the $xy$ plane. It is important that the
critical single well potential is harmonic (corresponding to hexapole
configuration of magnetic field), while either of the separated double
wells is quadrupole. We assume uniform $B_z$ is applied to the system.
In numerical calculations, we used the parameters close to the
experiment of Ref. \cite{Shin}: $d=150 \mu$m (this corresponds to the
 distance between the wires of 300$\mu$m), $B_{x_0}=24$G, $B_z(0)=1$G.


\subsection{Strong- and weak-field seeking states in two-wires trap}

The important feature of static magnetic traps is that it confine the
weak-field seeking state(s) (WFSS) which has a higher energy than the
strong-field seeking state(s) (SFSS). Therefore, there is some
difficulty in numerical preparation of initial states, since
straightforward imaginary time propagation would lead to SFSS. One need
to search for a solution using a WFSS ansatz which is derived from the
eigenvector of the $\cal{B}$ matrix. The eigenvalues of $\cal{B}$ are
$\pm B$ ( where $B=\sqrt{B_x^2+B_y^2+B_z^2}$) and 0 (the latter
corresponding to the neutral field-seeking state, NFSS), and the
eigenvector corresponding to WFSS is

\be
|-1>= \frac{1}{2 B}  \left(\begin{array}{c}
 (B - B_z)(B_x- i B_y)/B_{\perp} \\
 -\sqrt{2} B_{\perp}  \\
(B + B_z)(B_x + i B_y)/B_{\perp}
\end{array}\right)
\ee

Suppose we have configuration with  $\Delta B_x = 0$ (so the wells are
coalesced in a single harmonic well). When $(B_x \pm i B_y)/B_{\perp} =
e^{\mp 2i \phi}$ in the vicinity of the minimum of the well, where $
\phi$ is the polar angle around the point of the minimum. The
weak-field-seeking state order parameter has the form \be
\left(\begin{array}{c}
 \Psi_1 \\
\Psi_0 \\
\Psi_{-1}
\end{array}\right)
= \frac{1}{2 B}  \left(\begin{array}{c}
 (B - B_z)(B_x- i B_y)/B_{\perp} \\
 -\sqrt{2} B_{\perp}  \\
(B + B_z)(B_x + i B_y)/B_{\perp}
\end{array}\right) \psi e ^{i w \phi}=  \frac{1}{2 B}  \left(\begin{array}{c}
 (B - B_z) e ^{2 i \phi} \\
 -\sqrt{2} B_{\perp}  \\
(B + B_z) e^{-2i \phi}
\end{array}\right) \psi e ^{i w \phi}  , \label{WFSS}
\ee where $\psi$ is the amplitude (common for all three components). At
large $B_z$, almost all population is in $\Psi_{-1}$ component. In
order to avoid vorticity in this component, we should choose $w=2$.
Substituting ansatz (\ref{WFSS}) into GP equation, one obtains the
one-component equation for the amplitude $\psi$ which can be solved
using imaginary time propagation to find a ground state. However, it is
also possible to propagate the original three-component GP equation in
the imaginary time, restricting the solution to the anzatz
(\ref{WFSS}). We used both algorithms in our calculations.

One can see that one obtains a vortex with vorticity 4 from the initial
non-vortex state in the critical hexapole trap by reversal of $B_z$.
When $\Delta B_x \ne 0$, we have two quadrupole traps. In that case
reversal of $B_z$ leads to doubly quantized vortices in either trap, as
was studied in several works recently (for example, Ref.
\cite{Machida}). It is important that the component $\Psi_0$ in that
case contains singly quantized vortices.

\section{Vortex creation during realistic magnetic trap manipulations}

\subsection{Phase imprinting}

Bias fields $B_z$ and $B_x^B$ and the two-wires magnetic field $B^W$
are created by different  sources. During a switching-off process, they
can behave differently. Examples of nontrivial consequences  of
nonsynchronous decreasing processes of magnetic fields in other
situations were considered in Refs. \cite{Zhou,Zhang}.
In order to achieve topological imprinting, we consider here the
switching-off process with $B_z$ decaying faster than $B_{\perp}$ (in
our model calculations, we assume the fields $B_x^B$ and $B_{\perp}^W$
decay synchronously, while the field $B_z$ decays faster). We found a
time of the order of tens of $\mu$s is enough to transfer a large part
of the atoms to component $\Psi_0$ with singly quantized vortices in
each well.

In contrast, we found that in order for $B_z$ reversal to produce
considerable population of $\Psi_1$ component with doubly quantized
vortices, much longer times are required, i.e. hundreds of $\mu$s. In
the experiment of Ref. \cite{Shin} switching-off time was $T_{off}
\approx 20 \mu$s.  Larmor frequency at $B=1G$ is $\approx 700$ kHz
corresponding to $\tau_L=1.4 \mu$s. Therefore, despite very short
switching-off time, "adiabatic" phase imprinting still might take
place. We checked this guess numerically.

Four stages of the process were carefully examined: preparation of
initial state, splitting, switching-off process, and the free
expansion. Firstly, we prepare the initial state in the single
(critical) harmonic well using imaginary time propagation, as described
in the previous Section. (In the real experiment, the condensate was
prepared initially at $\Delta B_x=-140$mG in the bottom well of the
double-well potential, but it is not important for our present
purpose). When, the condensate was split by ramping $\Delta B_x$ from 0
to $100$ mG in real time. During this stage, almost all the population
is in $\Psi_{-1}$ component, because the $B_z$ field of 1G is large as
compared with magnetic field $B_{\perp}$ in the vicinity of the minima
of magnetic field, where the condensate resides. Provided the splitting
is slow, $\Psi_{-1}$ component remains without vortices (however, at
fast splitting it acquires phase winding, this case is considered in
the next subsection). During the third stage, the magnetic field was
turned off. Faster decay of $B_z$ may lead to creation of two singly
quantized vortices in component $\Psi_0$ in either of the two wells.
This would result in two fork-like structures in the interference
pattern of this component (only one of the forks points up, because
charges of the topological vortices in either well are the same
(positive)). The interference fringes were formed during the forth
stage of numerical calculations: expansion of the two condensates
without magnetic field. For numerical purposes, we also turned off the
gravity field in the forth stage, but we suppose this does not
influence results significantly. In the real experiment, typical time
of expansion was about 22 ms. In our calculations, we typically
calculate up to 4 ms only.

\begin{figure*}
\includegraphics[width=120mm]{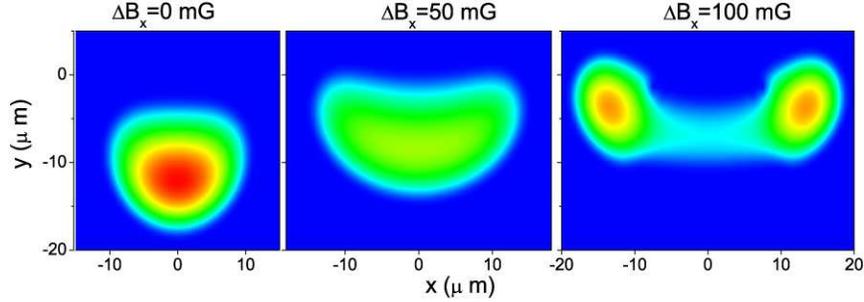}
\caption{Splitting of the condensate by ramping $\Delta B_x$ from 0 to
100 mG. Total density is shown. The origin of the coordinates is the
merge point without gravity. Parameters are close to the experiment of
Shin \emph{et. al};  splitting was fast: time of splitting about 5ms.}
\end{figure*}

In Fig.1, fast splitting of the condensate is shown. The condensate was
prepared in the merged well (at $\Delta B_x=0$). Splitting was done
approximately in 5 ms, much faster than in the experiment. In Fig.2,
slow splitting of the condensate is monitored (time of splitting about
30 ms). The final state is slightly different  about 2 \% of the
initial population was lost because during very slow splitting.  The
atoms that make transitions from WFSS to SFSS  fly away (they fly up
towards the wires and are removed by absorbing potential on the
boundaries of the mesh). The atoms that make transitions to NFSS, fly
down because of the gravity (and are also removed).

\begin{figure*}
\includegraphics[width=160mm]{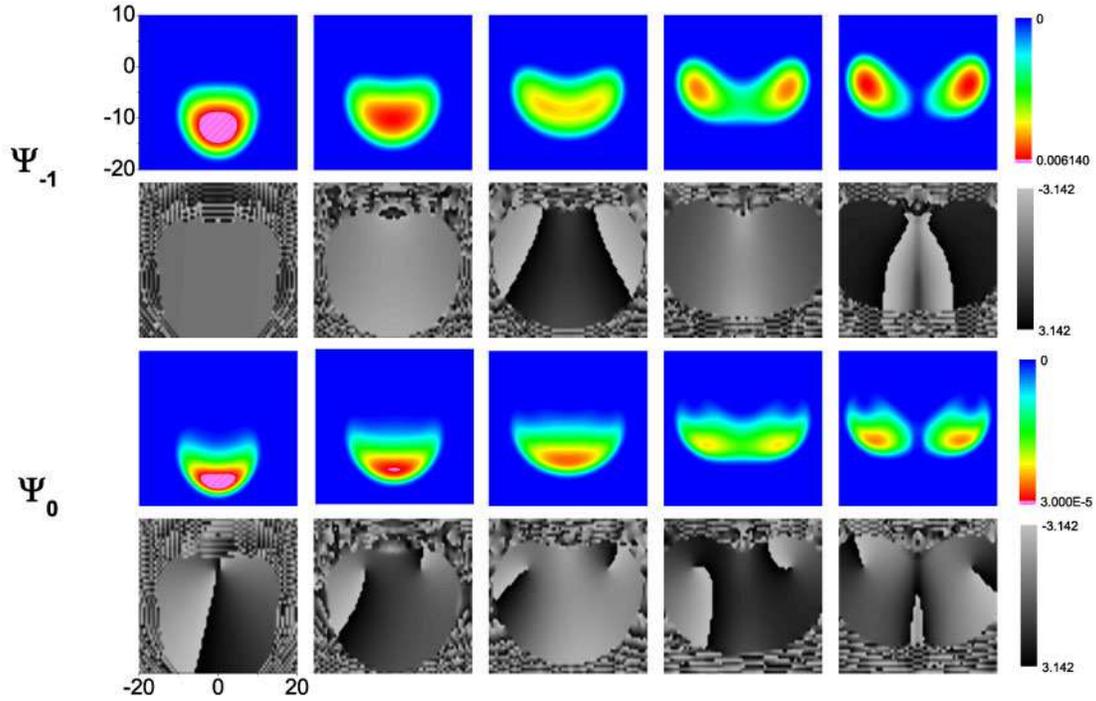}
\caption{Slow splitting of the condensate by ramping $\Delta B_x$ from
0 to 100 mG.  Time of splitting about 30ms. For each component, density
and phase profiles are depicted. In the component $\Psi_0$, singly
quantized  vortices of topological nature are seen. Their phase
singularities reside in the minima of the magnetic field which slowly
move as $\Delta B_x$ is increased. }
\end{figure*}

Densities and phases of the components of the split condensate are
presented. Component $\Psi_{-1}$ has no vorticity, while $\Psi_0$ has
singly quantized vortices in both wells  and $\Psi_1$ has doubly
quantized vortices. Since $B_z = 1G$, almost all population is in
$\Psi_{-1}$ component without vorticity.

\begin{figure*}
\includegraphics[width=120mm]{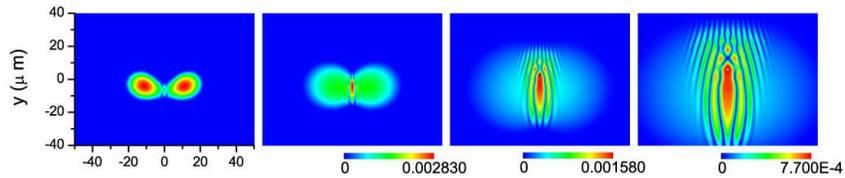}
\caption{Formation of fringes during expansion of the slowly split
condensates.  Switching-off of magnetic fields was done in such a way
that almost all population (more than 99\%) remains in $\Psi_{-1}$
component.}
\end{figure*}

In Fig. 3, interference fringes formed during expansion of the
condensates are shown. Magnetic fields were switched off in
approximately  $20 \mu$s in such a way that no transition from
$\Psi_{-1}$ component occured ($B_z$ was decreased more slowly than
$B_{\perp}$).

\begin{figure*}
\includegraphics[width=160mm]{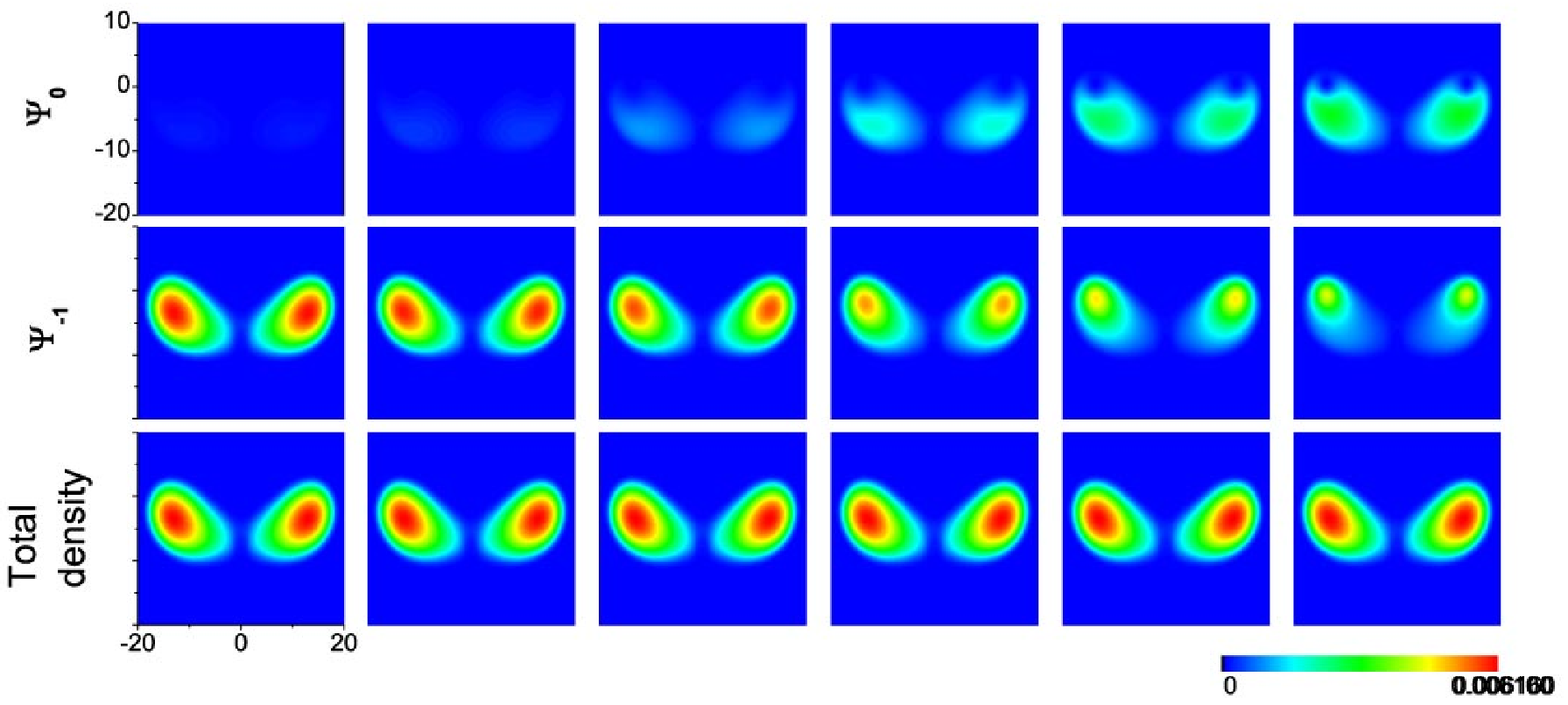}
\caption{Formation of singly quantized vortices in $\Psi_0$ component
during switching-off process. $B_z$ decay faster than $B_\perp$.  Time
of switching-off process is about 20 $\mu$s.  }
\end{figure*}

Then, we consider a case where $B_z$ is decreasing faster than
$B_{\perp}$. In Fig.4, the process of topological vortex formation is
shown during switching-off process. Magnetic fields were decreasing
exponentially, with $B_z$ decaying faster, and were turned off
completely at $t=0.06 \tau \approx 20 \mu s$ ($\tau \approx 350 \mu$s
is the characteristic time period used in the program). About half of
the total population were transferred to the component $\Psi_0$. An
important feature of nonadiabatic transitions during process of
switching-off can be noticed. The part of condensate in component
$\Psi_{-1}$ residing far from the minima of the magnetic field is more
easily converted to other components than the part near the minima. As
a result, decreasing $B_z$ slices out a part of the condensate around
the minima of the magnetic field and leaves it in the initial
$\Psi_{-1}$ component.  This is due to the fact that in the region
where $B_{\perp}$ is not zero, a nonvanishing gap between WFSS and NFSS
remains; besides, the stronger the field $B_{\perp}$ is, the more
slowly the total magnetic field rotates, making it easier for the spin
of an atom to follow it. Quantitatively, nonadiabatic transitions due
to nonsynchronously decreasing trapping magnetic fields (in other
configurations) were considered recently in Ref. \cite{Zhang} by
generalizing the Landau-Zener formula for the multilevel case. It can
be seen also that during fast switching-off process, the total density
is almost unaffected, therefore the nonlinear interaction coefficient
$g_n$ plays no role in the process.

\begin{figure*}
\includegraphics[width=100mm]{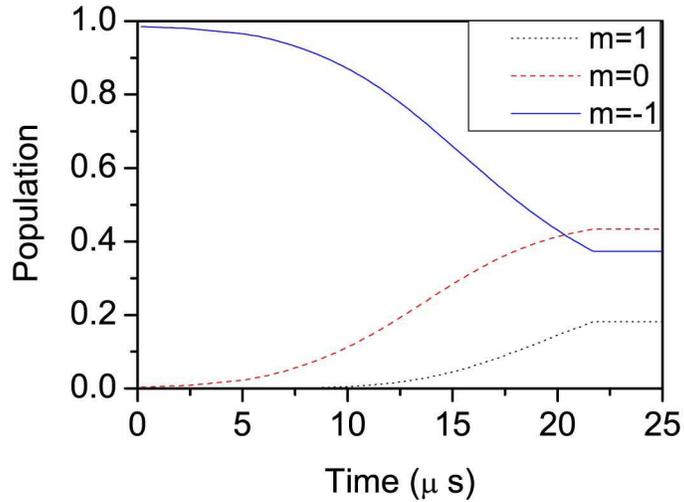}
\caption{Time evolution of populations of the components of the
condensate during switching-off. Time of switching-off process
$T_{off}$ is about 20 $\mu$s (at this time, all remaining magnetic
fields are abruptly turned off ). }
\end{figure*}

Time evolution of populations of the components is shown in Fig. 5. In
Fig. 6, interference fringes formed during expansion of the condensates
are shown. Fork-like structure is seen in the component $\Psi_0$.

\begin{figure*}
\includegraphics[width=140mm]{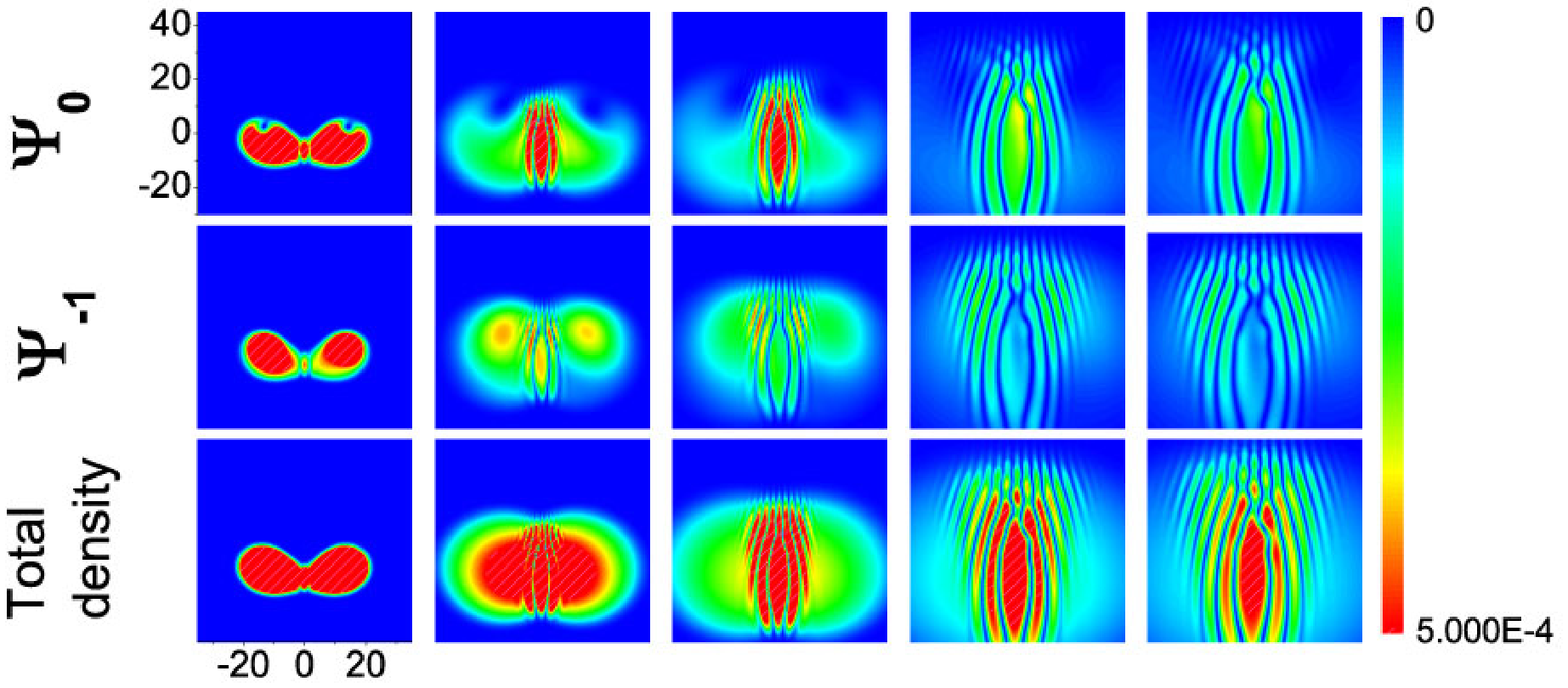}(a)
\includegraphics[width=80mm]{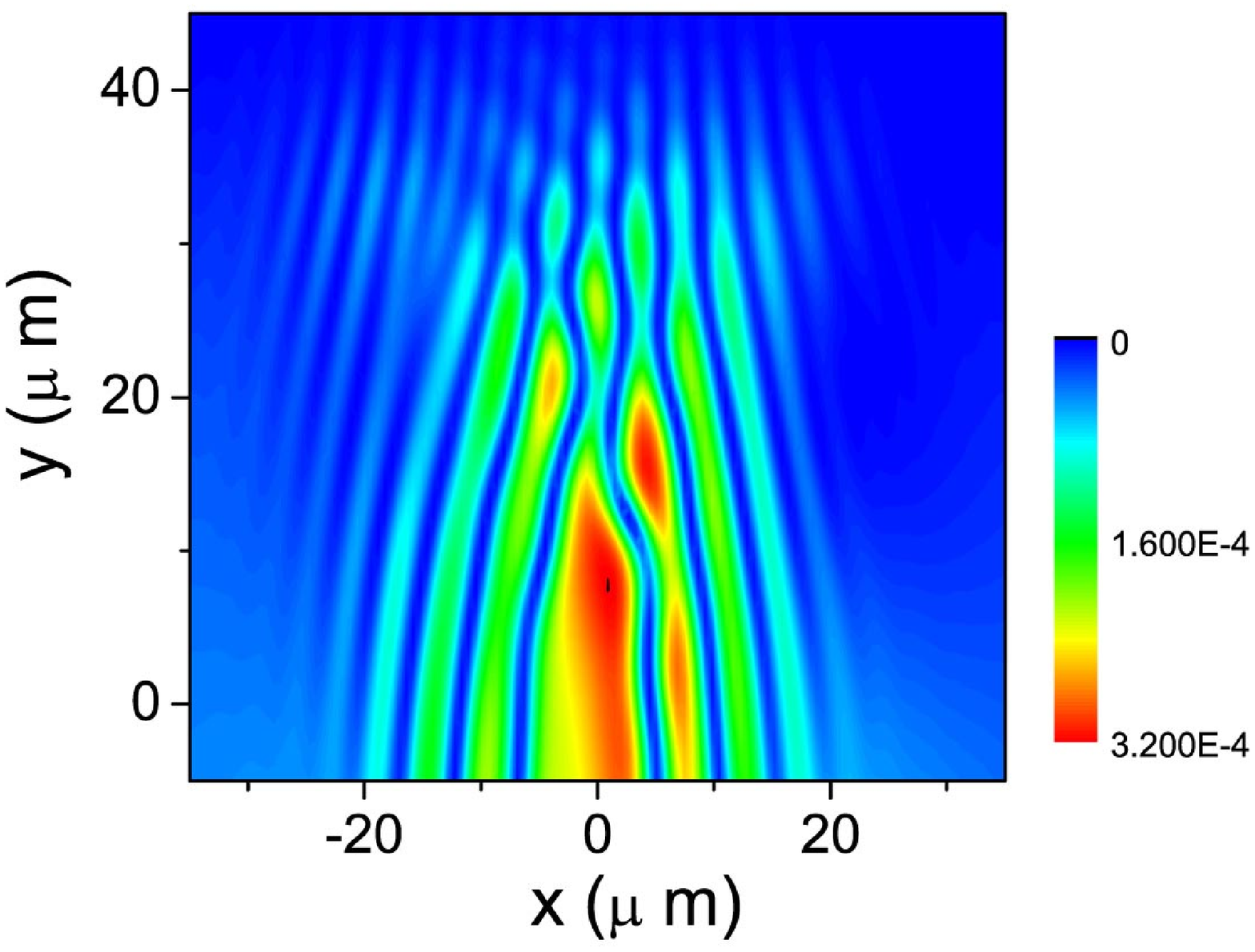}(b) \caption{Formation of fringes
during expansion of the condensate with imprinted vortices. Almost half
of the population were transferred to the component $\Psi_0$ during
switching-off. Topological vortices in component $\Psi_0$ has equal
charges in both wells. Only one fork points up in the $\Psi_0$
component. In total density pattern, the fork have not appeared yet.
However, note that in the experiment expansion took much longer time.
We see that the $\Psi_0$ component with vortices rotates clockwise
during expansion. After a long time, the fork therefore might appear in
the total density pattern too (our current computational recourses
forbid such a long-time propagation). (a) Density of components
$\Psi_0$,$\Psi_{-1},$ and total density, (b) Density of the component
$\Psi_0$ with the fork-like structure after approximately 4 ms of
expansion.
}
\end{figure*}

Topological vortices in component $\Psi_0$  in both wells has equal
charges. Therefore, only one fork points up (Ref. \cite{Ketterle3}).
The $\Psi_0$ component rotates clockwise during expansion due to equal
charge of the vortices, and the fork is moving up. Because of this,
although during our calculations it has not appeared in total density
profile, we believe that after a longer time it will show up (note that
almost half of the total population is in the $\Psi_0$ component). The
absence of the fork in the total density profile on early stages of
expansion is related to the above-mentioned behavior of the
switching-off process that leaves $\Psi_{-1}$ component residing near
the minima of magnetic field almost unaffected. After the switching-off
this component therefore is concentrated at the same places where the
phase singularities of component $\Psi_0$ are located.

For clarity, we give also an example of vortex formation without
gravity (Fig.9). Here, the both forks are clearly seen.

\subsection{Dynamical vortex creation}

\begin{figure*}
\includegraphics[width=160mm]{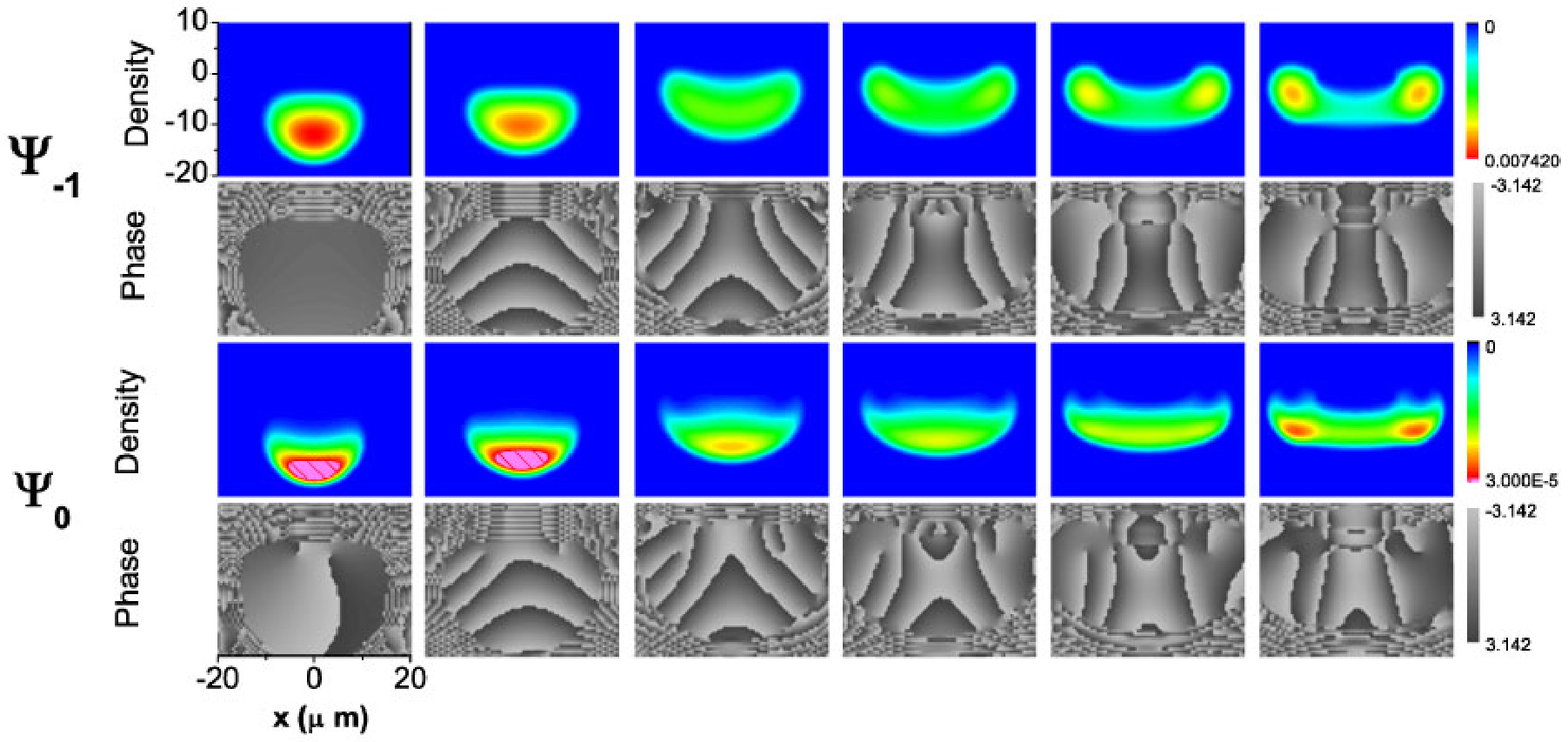}
\caption{Fast splitting of the condensate. Dynamical formation of
vortices: vortices are formed in all components (note that more than
99\% of total population is in $\Psi_{-1}$ component ). In the $\Psi_0$
component dynamical and topological vortices coexist. Topological
vortices are the same as in Fig.2.  }
\end{figure*}

In Fig.7, fast splitting of the condensate is monitored. Dynamically
created vortices appear in all the components (in component $\Psi_0$
dynamical  and topological vortices coexist). Centers of dynamically
created vortices (the phase singularities) lie outside the condensate,
but branch cuts go through the condensate causing phase winding across
it. So this situation is different from the so-called ghost vortices
appearing in studies of stirring condensates (Ref. \cite{Ueda}).
Although vortices are almost not visible in the density profile of the
condensate, its expansion leads to characteristic fork-like structures
in the interference fringes due to the phase winding.

\begin{figure*}
\includegraphics[width=160mm]{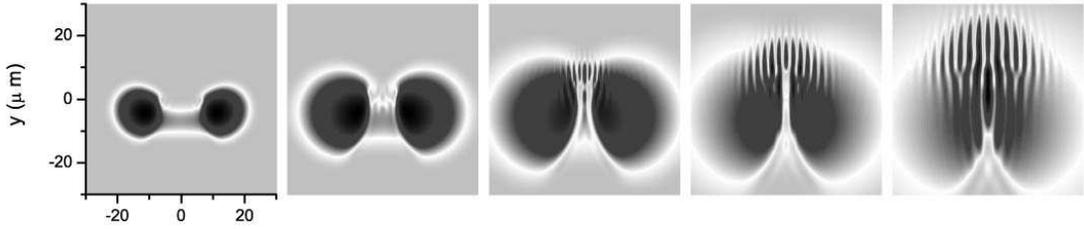}(a)
\includegraphics[width=80mm]{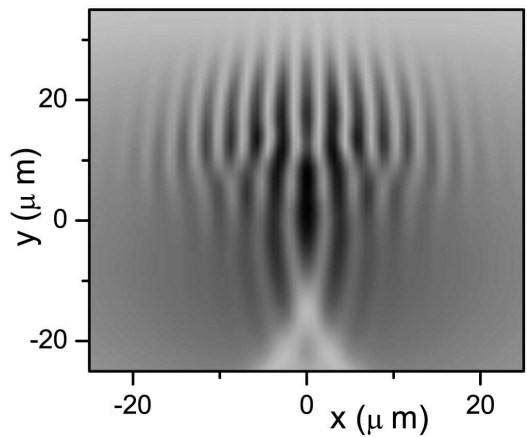}(b)
\caption{Formation of fringes during expansion of the rapidly split
condensates. During switching-off, almost no transfer occur to the
component $\Psi_0$. Dynamical vortices in the two wells have different
charges, therefore the two forks point up. }
\end{figure*}

Magnetic field was switched off in such a way that almost no transfer
to $\Psi_0$ occur. So, $\Psi_0$ component does not influence the
dynamics.  In Fig. 8, interference fringes resulted from the subsequent
expansion are shown. Two fork-like structures (which appear in all
components) are clearly seen. Vortices in the two condensates  have
opposite charges, therefore the forks point up (splitting of the
condensate on two parts in the presence of gravity effectively causes
rotation of each part in opposite directions).

\begin{figure*}
\includegraphics[width=160mm]{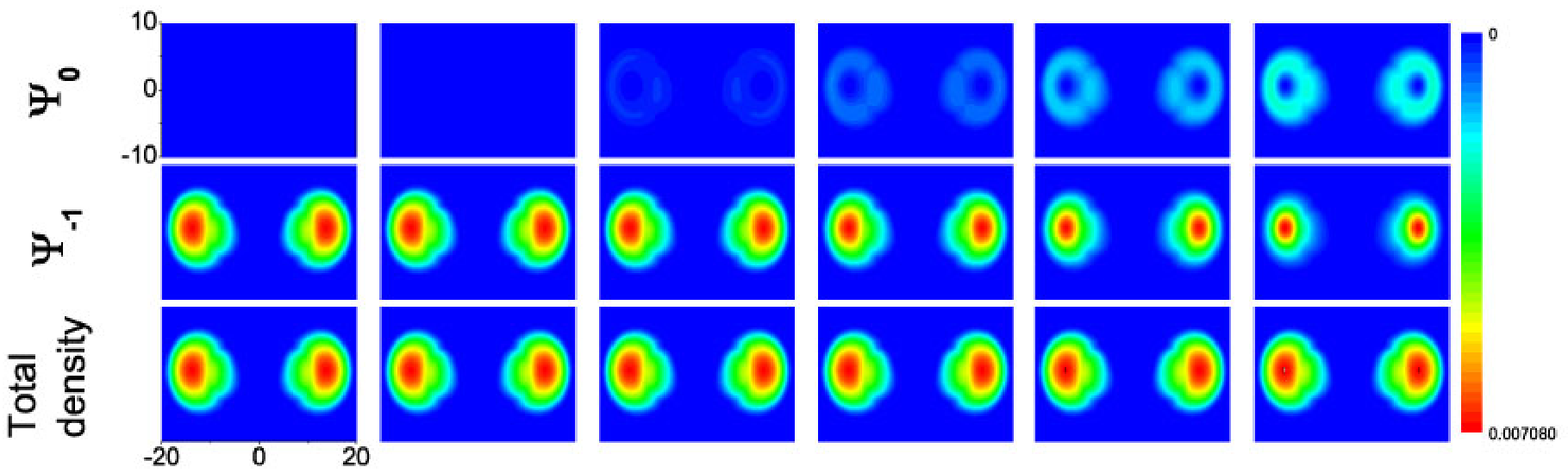}(a)
\includegraphics[width=100mm]{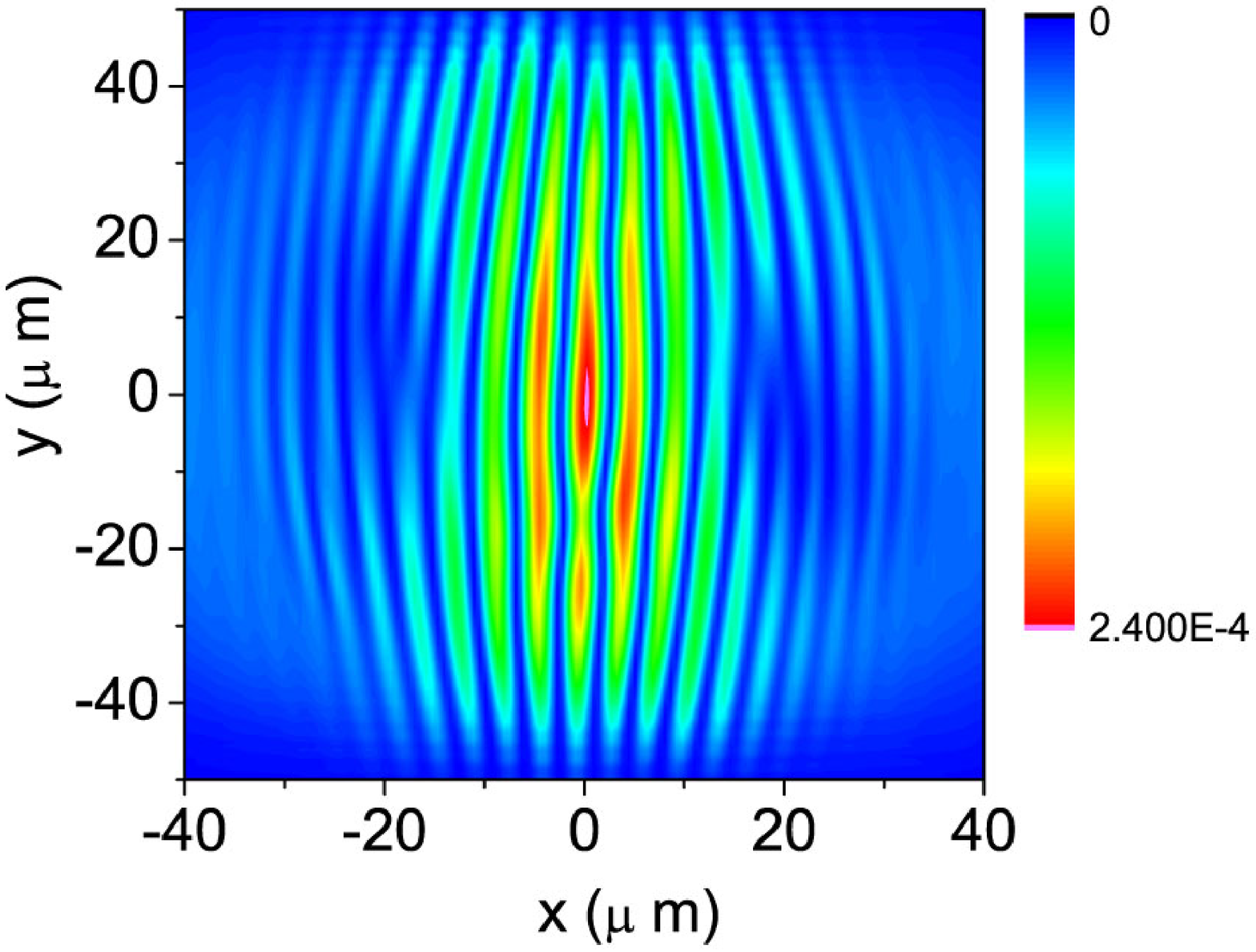}(b)
\caption{Dynamics without gravity. (a) Formation of singly quantized
vortices in $\Psi_0$ component during switching-off process. $B_z$
decay faster than $B_\perp$.  Time of switching-off process was about
20 $\mu$s. (b) Fork-like structures in the density of $\Psi_0$
component. }
\end{figure*}

To finish off, let us notice that the calculations reported were done
using the sin-DVR method (with a spatial mesh of $200 \times 160$ grid
points), and time propagation using the Runge-Kutta method (the method
is the generalization of the one used in Ref. \cite{hyro} to the case
of spinor condensates). We try to model the system as close to the
experimental one as possible. However, we found that, for example,
$g_s$ factor does not affect the results, i.e., the results are
qualitatively the same with $g_s=0$. The most important deviation of
the model from the experimental system is its reduced dimensionality.
The real system is three-dimensional, with nonuniform magnetic fields,
with an additional two pairs of wires creating magnetic fields to
compensate partly for nonuniformity of the base magnetic fields and
asymmetry, etc. Such complication is presently beyond our available
resources. However, we believe the study of the idealized model do
allow to discuss different mechanisms of vortex creation in the
experiment.

\section{Conclusion}

The authors of the experimental work \cite{Shin} supposed the phase
imprinting mechanism to be unlikely for explaining the appearance of
the fork-like structure and that ``the observed phase singularity
definitely shows the breakdown of adiabaticity''. To the contrary, we
found a realistic scenario based on nonsynchronous decreasing processes
of the magnetic fields can explain the phase singularity even within
the assumption of adiabatic evolution.

However, we found that fast splitting can also lead to dynamical vortex
creation, and that the dynamically created vortices produce
interference patterns with the forks of better contrast. Detailed study
of these processes is left for future research. In any event, the
splitting of a spinor BEC is a rather violent process so that further
consideration of atom interferometry with spinor BEC is necessary.

\section{Acknowledgements}

A.P. Itin is supported by JSPS.  This work was also supported in part
by Grants-in-Aid for Scientific Research No. 15540381 and 16-04315 from
the Ministry of Education, Culture, Sports, Science and Technology,
Japan, and also in part by the 21st Century COE program on ``Coherent
Optical Science''. TM was also supported in part by a financial aid
from the Matsuo Foundation. We would like to acknowledge useful
discussions with Prof. Y.S. Kivshar and Prof. Nakagawa.

\end{document}